# A Simulation Study on the Feasibility of Radio Enhancement Therapy with Calcium Tungstate and Hafnium Oxide Nanoparticles

Nicholas J. Sherck,[1,3] and You-Yeon Won[1,2,3*]

[1] School of Chemical Engineering, Purdue University, West Lafayette, Indiana 47907
[2] Purdue University Center for Cancer Research, West Lafayette, Indiana 47907
[3] LoDos Theranostics, West Lafayette, Indiana 47906

ABSTRACT: Herein is a simulation study on the radio enhancement potential of calcium tungstate ($CaWO_4$) and hafnium oxide ($HfO_2$) nanoparticles (NPs) relative to gold (Au) NPs. The work utilizes the extensively studied Au NP as the "gold standard" to which the novel materials can be compared. All three materials were investigated *in-silico* with the software Penetration and Energy Loss of Positrons and Electrons (PENELOPE) developed by Francesc Salvat and distributed in the United States by the Radiation Safety Information Computational Center (RSICC) at Oak Ridge National Laboratory. The aims are: (1) Do $CaWO_4$ and $HfO_2$ NPs function like Au?, and (2) if not, how else might they function to enhance radio therapy? Our investigations have found that $HfO_2$ likely functions as Au, but not as effectively. $CaWO_4$ likely does not function as Au, and we propose that $CaWO_4$ may exhibit cancer killing traits through its intrinsic UV luminescence property.

## 1. Introduction

Gold nanoparticles have been the premier material investigated for radiation sensitization since 2000 when they were first injected into tumors by Herold *et al.*, and was further investigated by Hainfeld *et al.* in 2004 (Herold *et al.*, 2000; Hainfeld *et al.*, 2004). To date there have been several hundred articles published on radiation sensitization by Au-NPs reporting macroscopic dose enhancement factors of 10 to 100 percent, but there remains much uncertainty as to the actual tumoricidal mechanism(s) (Schuemann *et al.*, 2016). Butterworth and colleagues have reviewed several pathways through which NPs can stress the cell, such as: (1) localizing ionizing radiation in the vicinity of the NP directly damaging DNA, (2) indirect generation of reactive oxygen species (ROS) causing oxidative stress, and (3) cellular induced stress due to the presence of nanoparticles (Butterworth *et al.*, 2012). In 2011, McMahon and colleagues demonstrated the importance of not assessing Au NP dose enhancement on a macroscopic scale, but rather at nanometer depths from the surface; they reported doses in the hundreds to thousands of Gy at the surface of the NP (McMahon *et al.*, 2011). They implemented the Local Effect Model (LEM), and demonstrated the accurate prediction of cell survival fraction after radiation therapy enhanced with Au NPs (Tommasino *et al.*, 2013; Elsasser and Scholz, 2007). The findings by McMahon indicate that a key metric in evaluating novel materials is to evaluate their local dose enhancement. This is the method we use to address whether $CaWO_4$ and $HfO_2$ NPs are likely to function like Au.

---

[*] To whom correspondence should be addressed. E-mail: yywon@ecn.purdue.edu

CaWO$_4$ is a luminescent material, that was first reported on by Thomas Edison in the Journal Nature in 1896, and is today being used in Dark Matter search experiments due to its ability to convert high energy particles into light through a scintillation process (Sivers *et al.*, 2012). The scintillation property of bulk CaWO$_4$, peak emission typically at 420 nm, has been thoroughly documented: Moszyński *et al.* characterized the process at both room temperature and liquid nitrogen temperatures (Moszyński *et al.*, 2005), Mikhailik and colleagues investigated the scintillation process under different excitation energies ≤ 35 keV (Mikhailik *et al.*, 2005), Kraus *et al.* investigated the difference between ZnWO$_4$ and CaWO$_4$ (Kraus *et al.*, 2005), Sivers and coworkers examined the effect of annealing on the optical and scintillation properties, Bailiff and colleagues addressed the one- and two-photon luminescence and band-gap of CaWO$_4$ crystals (Mikhailik *et al.*, 2004), the recombination process of electron-hole (e-h) pairs was investigated by Nagirnyi *et al.* *(Nagirnyi et al., 1998)*, and finally the thermal quenching of e-h pairs has been discussed by Mallory and coworkers (Beard *et al.*, 1962). These investigations were for bulk CaWO$_4$ crystals and almost solely driven over the past decade by the search for Weakly Interacting Massive Particles (WIMP) in physics. There have also been a small number of investigations by groups on the luminescence of CaWO$_4$ at the micro- and nanocrystal scale in contrast to the bulk scale (Su *et al.*, 2008; Cavalcante *et al.*, 2012; Basu *et al.*, 2014). Our group has recently described a synthesis protocol and formulation to encapsulate CaWO$_4$ NP in a bioinert block copolymer (BCP) micelle, while ensuring the fidelity of these encapsulated-NP luminescence under excitation (Lee *et al.*, 2016). Lee *et al.* measured the fluorescent peak at 420 nm (2.9 eV), with significant emission < 400 nm into the UV spectrum.

The second material of interest is that of HfO$_2$ being developed by Nanobiotix. In 2012 Maggiorella *et al.* reported on hafnium oxide's (HfO$_2$) potential to enhance radiation therapy supported by simulation, clonogenic assays and in a mouse model (Maggiorella *et al.*, 2012). HfO$_2$ is a scintillating material like CaWO$_4$ emitting at 4.2 – 4.4 eV at 10 K, but is thermally quenched by a factor of more than 80× at room temperature (Kirm *et al.*, 2005). Thus, even though HfO$_2$ emits strongly in the UV range, the effect is not appreciable at body temperature.

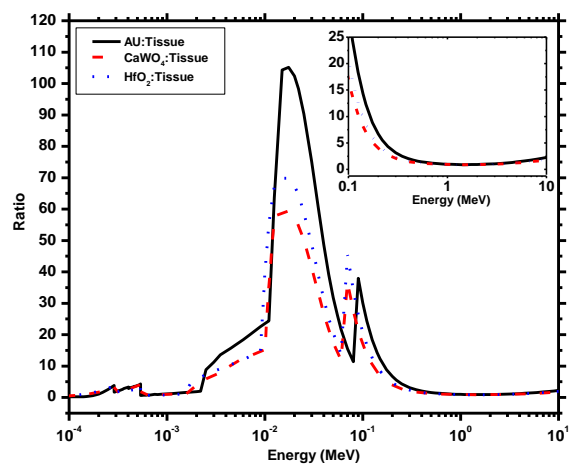

**Figure 1:** Comparison of the ratios of the photon MAC for Au, CaWO$_2$ and HfO$_2$ relative to the International Commission on Radiation Units (IRCU) four-component soft tissue composition as a function of photon energy. Figure generated from PENELOPE's databases (Salvat, 2015).

This paper seeks to provide a concise comparison of the three materials Au, $HfO_2$ and $CaWO_4$ for radiation therapy to assess their potential relative to Au. To start, one can compare the ratios of the photon mass attenuation coefficients (pMAC) of the three different materials relative to that of soft tissue, Figure 1; from this figure one would expect the radiation enhancement of the materials to follow the order Au > $HfO_2$ > $CaWO_4$. The largest therapeutic benefit would be expected at energies around 10 keV photon irradiation, but many studies suggest a significant enhancement beyond what is expected by theory up to MeV energies for Au as was pointedly reviewed by Butterworth *et al.* (Butterworth *et al.*, 2012). Butterworth *et al.*'s findings are a forewarning to proceed with caution when attempting to translate theoretical and simulation predictions to complex biological processes.

## 2. Methods

### 2.1 *General*

The simulations conducted in this paper were conducted with the Monte Carlo based software package: Penetration and Energy Loss of Positrons and Electrons (PENELOPE) developed by Francesc Salvat and distributed in the United States by the Radiation Safety Information Computational Center (RSICC) at Oak Ridge National Laboratory (Salvat, 2015). The main steering program for the PENELOPE physics model was that of PenEasy developed by Brualla and coworkers (Sempau *et al.*, 2011). The steering program (also known as the main program) is used to define specific tallies of interests, e.g., dose distribution or emission spectrums. The cutoff absorption values were set to 100 eV for both electrons and positrons, and 50 eV for photons. The cutoff absorption values are used to decide when to stop following primary or secondary particles or photons whose energies have fallen to a certain level. The simulation parameters C1, C2, WCC and WCR were set to 0.1, 0.1, 10 and 10 respectively, please see the material from Salvat on the PENELOPE model for more details on the meaning of these parameters (Salvat and Fernández-Varea, 2009; Sempau *et al.*, 2003; Salvat, 2015). Maximum step sizes for the program were set to $< 1/10^{th}$ the overall thickness of the material. Throughout the paper many units are given for a quantity per history. The history refers to the incident high energy photon beam, $\gamma$ photon, simulated. Each simulation may simulate up to $10^5 - 10^{10}$ total incident $\gamma$ photons, and each one is referred to as a history. Most $\gamma$ photons will not interact with a particle, but still count to the average, as happens in a reality.

### 2.2 *Simulation Geometry 1*

The first geometry was comprised of a spherical particle in vacuum encapsulated in a detector. This setup allows for an isolated comparison of the material properties, and a closer look at the role size plays in the intrinsic absorption of incident radiation energy by the NPs. All three materials were simulated at NP sizes of 3, 10, 50, 100, 200 and 500 nm diameters at monotonic, photon beam energies incident on their cross-sectional areas of 30, 160, 1200 ($CaWO_4$ only) and 6000 keV. Energy tallies were used to determine the incident energy that was deposited in either the NP or the detector in units of electron volts per simulated incident photon (eV/hist.). This value was also normalized to the mass of the NP (eV/g/hist.). The NP in vacuum is an idealization of what is expected to be occurring in medium, where there would be potential for radiation to enter the particle from all sides due to scattering processes in the medium and for photon beam attenuation. The spectrum of electrons/photons generated and emitted from the NPs was recorded, along with a quantitative analysis using two averages of the spectrums given by equations 1 and 2:

$$eqn. (1) \quad \langle Energy \rangle = \frac{\sum_{i=1}^{B} n_i * E_i}{\sum_{i=1}^{B} n_i};$$

$$eqn. (2) \quad \langle Count \rangle = \frac{\sum_{i=1}^{B} E_i * n_i}{\sum_{i=1}^{B} E_i}$$

where B is the total number of bins in the histogram over the energy spectrum, $n_i$ is the counts of particles emitted with energy in the $i^{th}$ energy bin, and $E_i$ is the average energy of the $i^{th}$ bin.

2.3 *Simulation Geometry 2*

The second geometry was used to analyze the effect of irradiation in a medium and the radial dose distribution around the NP. The geometry involves a cylindrical object filled with a medium and one NP of varying size centered at x = y = z = 0. The medium is water with a mass concentration of 1 mg per gram water of the respective NP material uniformly dispersed. Figure 2 is a representative schematic of geometry 2:

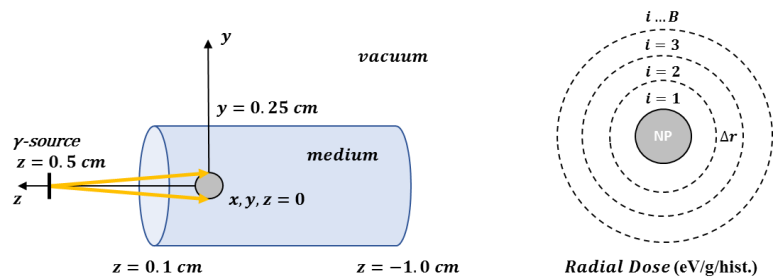

**Figure 2:** Schematic representation of geometry 2. The figure is not to scale. Everything outside the medium is a vacuum. The *i* and B do not correspond to eqns. 1 and 2, but rather 3 and 4.

The geometry is used to analyze the effect of a photon beam traveling through a medium as well as the dose enhancement factor (DEF) around the NP. The macroscopic enhancement factor is well-known not to be in good agreement with experimental results, but rather the Local Enhancement Model (LEM) should be used (McMahon *et al.*, 2011; Scholz and Kraft, 1994; Lechtman *et al.*, 2013). Using the LEM it has been shown that experimentally measured cell survival curves can be calculated for Au NPs. Equations 3 and 4 were used to calculate the local dose enhance factor (DEF$_{loc}$) for a given NP material, size, irradiation energy and depth in medium:

$$eqn. (3) \quad DEF_{loc,i} = \frac{D_i^{NP}}{D_i^{Base}};$$

$$eqn. (4) \quad \langle DEF_{loc} \rangle = \frac{\sum_{i=1}^{B} DEF_{loc,i} * \Delta v_i}{\sum_{i=1}^{B} \Delta v_i}.$$

The local DEF, $V_{loc}$, can be averaged over the entire volume, $V_T$, with a certain number density, $\hat{n}$, of particles and added to the baseline absorption to determine the radiation therapy enhancement with nanoparticles, equation 5, and can be related to cell survival curves based on the semi-empirical, linear-quadratic (LQ) model, equation 6 (Barendsen, 1997; Franken *et al.*, 2006; Brenner, 2008):

$$eqn.\,(5) \quad D_{NP} = \frac{D_{Base} * \langle DEF_{loc} \rangle * \hat{n} * V_T * V_{loc}}{V_T} = \frac{D_{Base} * \langle DEF_{loc} \rangle * m_T * V_{loc}}{V_T * \frac{4}{3}\pi R_p^3 * \rho_p}$$

$$where \quad D_T = D_{NP} + D_{Base} \;(in\ units\ of\ Gy);$$

$$eqn.\,(6) \quad S = e^{-(\alpha D_T + \beta D_T^2)}$$

$$where\ \alpha\ and\ \beta\ are\ specific\ to\ a\ cell\ line.$$

## 3. Results

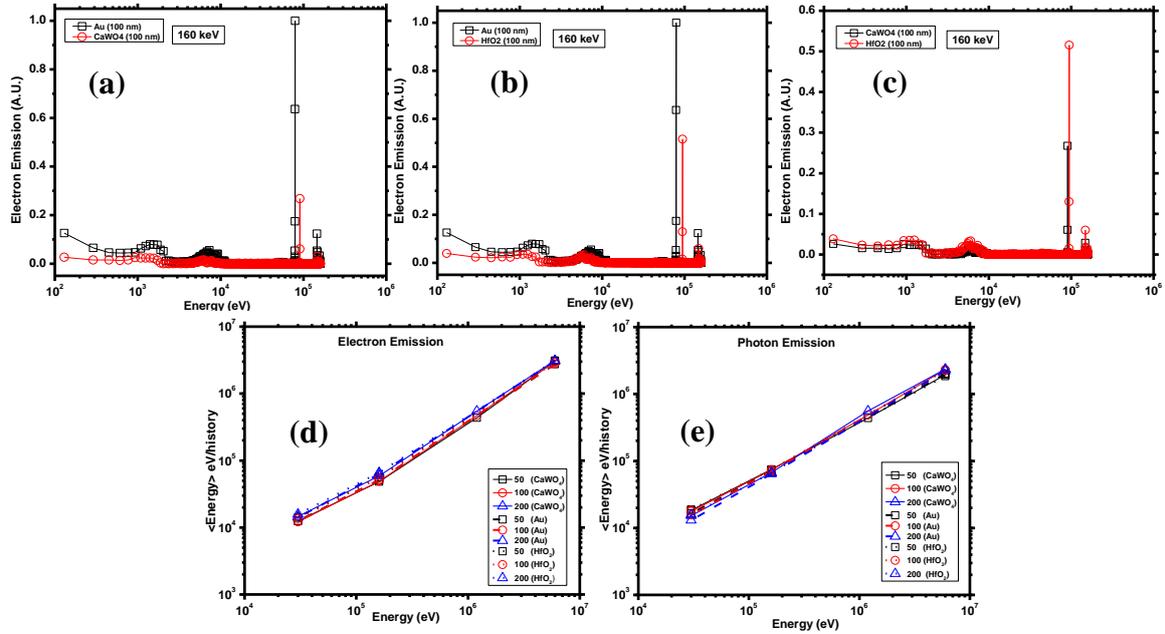

**Figure 3:** The electron generation spectrum for 100 nm NPs irradiated with a monotonic $\gamma$ photon beam. The energy spectrums are normalized to Au. (a) 160 keV, Au and $CaWO_4$ electron emission. (b) 160 keV, Au and $HfO_2$ electron emission. (c) 160 keV, $CaWO_4$ and $HfO_2$ electron emission. (d) Au, $CaWO_4$ and $HfO_2$ average electron energy emission spectrums, eqn. 1. (e) Au, $CaWO_4$ and $HfO_2$ average photon energy emission spectrums, eqn. 1.

Using geometry 1, a few representative plots of the electron emission spectrums from the NPs, Figures 3(a), 3(b) and 3(c), were produced. Emission spectrum plots are useful for qualitative analysis between two materials, but are not ideal for quantitative analysis between materials. Therefore, Figures 3(d) and 3(e) display the average energy as given by equation 1 above. It can be discerned that the average energy for electrons and photons for Au, $CaWO_4$ and $HfO_2$ are similar over a range of photon irradiation energies. This is due to their similar electronic structures, atomic numbers of 79, 74, 72 for Au, W and Hf, respectively. When the count is weighted by energy, the number of electrons and photons emitted is less following the order Au > $HfO_2$ > $CaWO_4$ over varying photon irradiation intensities, see the SI material section 1, and is as expected from the pMAC in Figure 1.

The first geometry was also used to evaluate the absorption by a NP of a certain size and irradiation for luminous intensity calculations for $CaWO_4$ NPs. Figure 4(a) displays the absorption results normalized to particle mass. From this result it is expected that there is no correlation between particle size and the energy absorption per particle mass. This would indicate that at a given macroscopic mass concentration in a given volume, the particles will absorb the same amount of radiation regardless. In the case of Au NPs, this energy is wasted energy, because this energy is not redistributed outside of the particle.

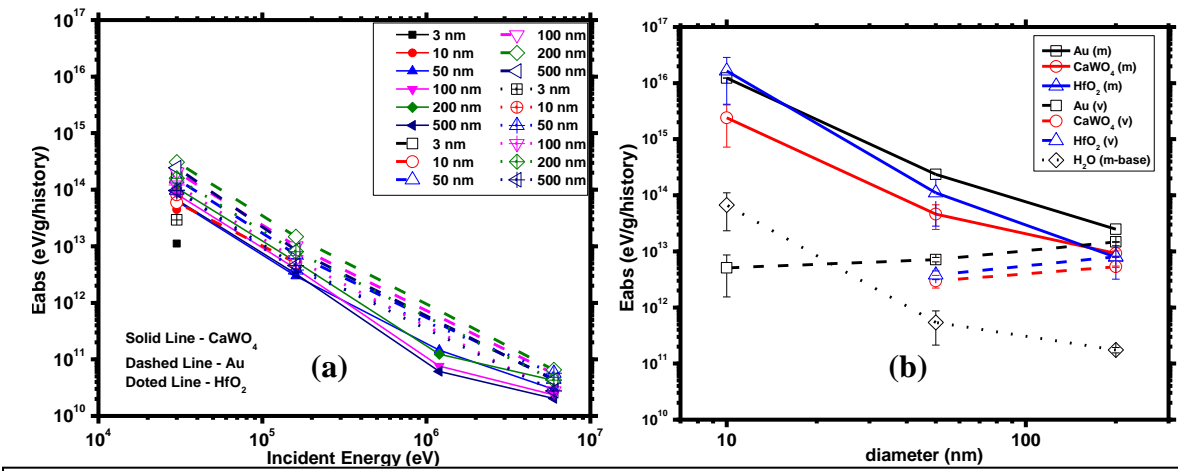

**Figure 4:** The absorbance of energy intrinsically by the NP for varying materials, sizes and photon irradiation energies. (a) Absorbance for particles in geometry 1, normalized to particle mass. (b) Intrinsic NP absorbance for 160 keV irradiation at different sizes for both geometry 1 and 2. In the legend, (m) denotes simulations in medium and (v) denotes simulations in vacuum. Base denotes a NP filled with water. Normalized to particle mass, error bars are one standard deviation n = 3 for each data point.

The second geometry is a better realization of radiation enhancement by nanoparticles dispersed in a medium or inside of cells, but not for intrinsic energy absorption by the NPs; Figure 4(b) is a plot of the intrinsic NP absorption at 160 keV $\gamma$ photon irradiation using geometry 2. From Figure 4(b) a trend appears on NP size and the amount of energy absorbed below 200 nm NP diameters. This suggest that NPs of smaller diameter absorb more energy per mass per history. After further investigation it is believed that this is an artifact of the simulation, because NPs comprised of water also have an increasing absorption per mass as they decrease in size ($H_2O$ base case in Figure 4(b)).

As discussed earlier, an important metric in appraising new materials for radiation sensitization is to compare the local DEF at the NP surface. It is well known that Au NPs are increasingly effective at lower photon irradiation energies – in the tens of keV region (Chithrani *et al.*, 2010). It is also understood that high energy, photon beams become attenuated as they travel through the medium. Given these facts, it was decided to investigate the NP DEF only at 160 keV and 6 MeV rigorously, see geometry 2. Of course the magnitude of these trends will change with depth into the medium due to a decrease in the number of incident photons, but relative performance should not change between the materials. Using equation 4, a local DEF can be computed, averaged over different distances from the surface of the NPs. The results for 160 keV irradiation are presented in Figures 5(a), 5(b) and 5(c), and the results for 6 MeV are in SI Section 2. The local DEFs followed the expected trend, Au > $HfO_2$ > $CaWO_4$. The key finding is that Au delivers a significantly higher dose (4 – 100×) compared to $CaWO_4$. The difference is not as stark for $HfO_2$, but becomes noticeable at the larger NP sizes. The peaks are due to the tradeoffs between the increased interaction probabilities of a larger particle with incident radiation, and a decrease in the probability of escape for a generated electron/photon at a given energy.

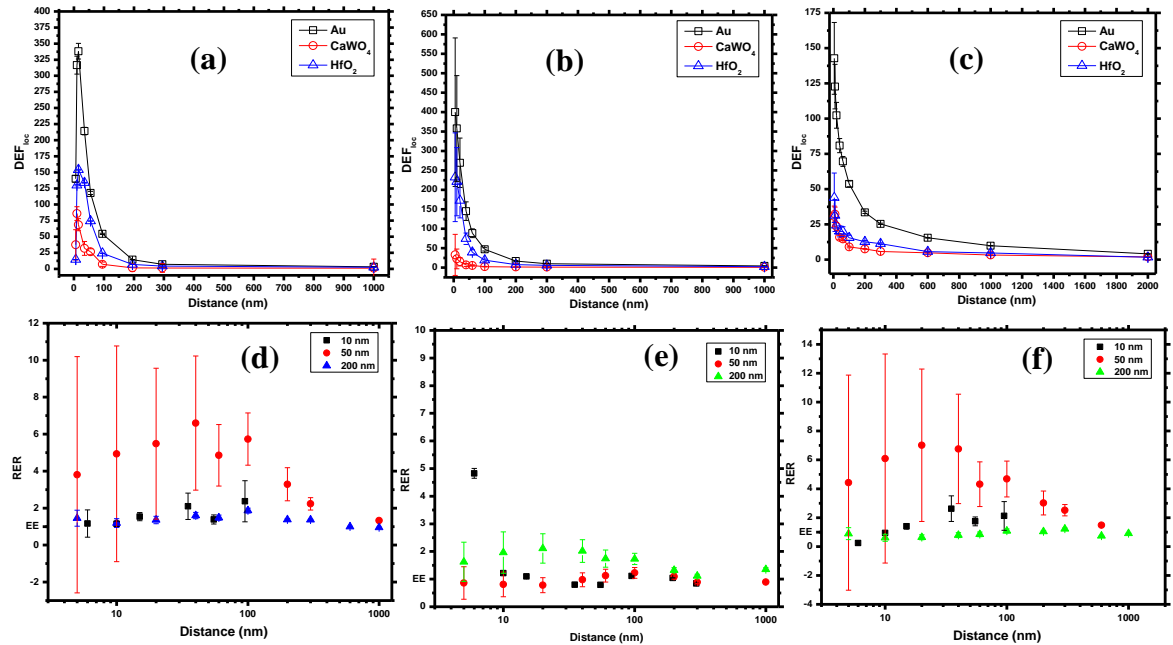

**Figure 5:** **(Top Row)** Plots of the local dose enhancement factor ($DEF_{loc}$) calculated at different distances from the surface of the NP (x-axis) for a monotonic, 160 keV photon beam. (a) 10 nm NP. (b) 50 nm NP. (c) 200 nm NP. The error bars are ± 1 standard deviation (SD) calculated from three MC simulations for each material against a baseline water medium under similar irradiation. **(Bottom Row)** Plots of the RER for (d) Au:$CaWO_4$, (e) Au:$HfO_2$, (f) HfO2:CaWO4 at varying distances from the surface. See SI Section 2 for 6

Another way to look at the local DEF on the macroscopic level is by using equation 5, which takes into account the overall NP dose enhancement's ($D_{NP}$) dependence on the size and density of the NP. Taking a ratio of equation 5 for two materials at constant mass concentration and NP size yields a relative effectiveness ratio (RER), equation 7:

$$eqn. (7) \quad RER = \frac{\rho_p^2}{\rho_p^1} \frac{\langle DEF_{loc} \rangle^1}{\langle DEF_{loc} \rangle^2} = \varepsilon \frac{\langle DEF_{loc} \rangle^1}{\langle DEF_{loc} \rangle^2}.$$

Equation 7 shows that even a particle with a larger local DEF may be inferior to a particle with a lower DEF, but with a much lower density than that of the other material. For Au:$CaWO_4$, Au:$HfO_2$ and HfO2:CaWO4, ε is 3.2, 2.0 and 1.6, respectively. This indicates that the ratio of the local DEFs should be greater than ε for a given material to be a better option than another. This is of course only true for particles of the same size and at the same mass concentration. In actuality, for *in vivo* experiments, the mass concentration inside a cell is likely not to be the same for NPs of the same size, but with different densities. This is due to the fact that size determines the cellular uptake of the material; that means that a NP of a given size will have the same number concentration inside a cell, and therefore not the same mass concentration. One should heed this finding when evaluating particles in *in vitro* where often mass concentrations are held to be the same. Figures 5(d), 5(e) and 5(f) display the RER for Au:CaWO$_4$, Au:HfO$_2$ and HfO2:CaWO4, respectively. One can conclude that CaWO$_4$ likely will not perform to the level of Au, given the lower local DEFs and RERs. The results indicate HfO$_2$ does not perform as well as Au as a radio therapy enhancer, but could still be expected to exhibit a therapeutic effect.

## 4. Discussion

The simulation results are believed to be reasonable, especially for relative comparisons between materials, as other simulations and experiments have shown local DEFs to be on the same order of magnitude (Lee *et al.*, 2012; Jones *et al.*, 2010; Regulla *et al.*, 1998). Our results have indicated that CaWO$_4$ should not perform at the same level as Au determined from the local DEF-LEM model. On the contrary, HfO$_2$ should still yield a therapeutic effect, albeit at a reduced efficacy to that of Au. HfO$_2$ is in clinical trials, but it is not clear if there has ever been a head-to-head comparison of Au and HfO$_2$. Likewise, there have been no direct comparisons of Au with CaWO$_2$. If CaWO$_2$ exhibits a therapeutic effect, it would be expected from these findings to proceed through a separate mechanism. This study is only considering the physical enhancement from the NPs, but as some have pointed out, there are likely other mechanisms at play, e.g., chemical enhancement by the NPs themselves (Lee *et al.*, 2012; Sicard-Roselli *et al.*, 2014).

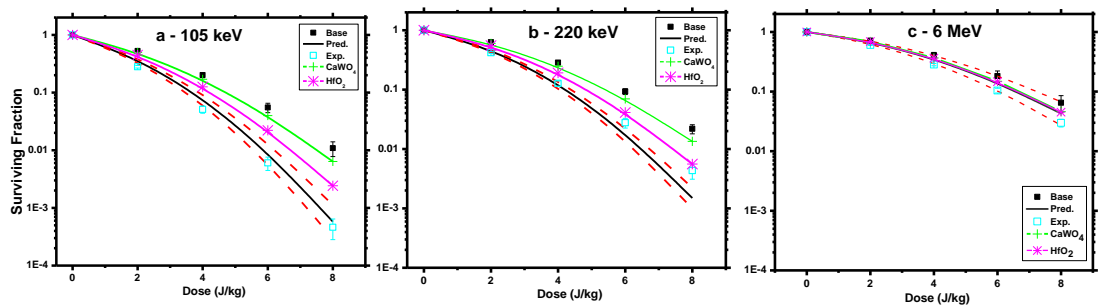

**Figure 6**: Comparison of simulation to experimental cell surviving fraction (HeLa) results from literature for 50 nm diameter Au NPs. Displayed are also the predicted survival curves with CaWO$_4$ and HfO$_2$, green and magenta lines, respectively. The local DEF was calculated at 1000 nm from the NP surface. (a) Simulation conducted at 160 keV $\gamma$ photon irradiation. Experiment conducted at 105 keV. (b) Simulation conducted at 160 keV $\gamma$ photon irradiation. Experiment conducted at 220 keV. (c) Simulation conducted at 6 MeV $\gamma$ photon irradiation. Experiment conducted at 6 MeV. All error bars are $\pm$ 2 SDs. Red dash lines are $\pm$ 2 SDs for the predicted survival curves. The $\alpha$ and $\beta$ values used, equations 5 and 6, are from the Chithrani *et al.* (Chithrani et al., 2010): (a) Base/Predicted – 0.237 and 0.041, Experimental – 0.528 and 0.054, (b) Base/Predicted – 0.150 and 0.041, Experimental – 0.352 and 0.041, and (c) Base/Predicted – 0.110 and 0.029, Experimental – 0.191 and 0.031.

The simulation was further used to predict experimental cell survival curves under $\gamma$ photon irradiation at different energies. The simulation was conducted at 160 keV, while the experiments were conducted above and below at 105 and 220 keV. As expected, the model should either under (105 keV) or over (220 keV) predict, because at lower irradiation energies, the enhancement is increased. As expected, in Figures 6(a) and 6(b) this trend is observed. The experiment and simulations were both conducted at 6 MeV, Figure 6(c). Both experiment and simulation with Au NPs are in agreement with a significantly lower enhancement in the survival curve at 6 MeV. The model's agreement with experimental trends of cell survival curves, lends confidence in the model. Conversely, the model's correlation with experiment again further validates the LEM theory for radiation therapy.

If $CaWO_4$ does not function as Au, then perhaps the obvious candidate mechanism, in the case of $CaWO_4$ NPs, is its UV emission under irradiance. Just this year Varghese and coworkers have shown that pretreatment of $CaWO_4$ material by irradiation with a high energy, electron beam can significantly shift the peak emission from 420 nm down to at least around 340 nm, well within the UV-B spectrum (Aloysius Sabu *et al.*, 2016). They have also indicated that this shift is accompanied with no loss in spectral emission intensity. With a shift to the 340 nm range, accompanied by a broad tail into the UV-C spectrum, it is expected to further enhance the tumoricidial effect, as it is known that UV-C, UV-B and even UV-A radiation can cause DNA damage and cell death through both a direct and an indirect means (Miwa *et al.*, 2013; Godar, 1999; Kielbassa *et al.*, 1997).

### Table 1: Scintillation Properties of $CaWO_4$

| Quantity | nano | Micro | macro | References |
|---|---|---|---|---|
| $E_{gap}$ (eV) | 5.5[e,b] | n.d. | 5.2-4.6[b] | [a](Cavalcante *et al.*, 2012) |
| $E_{photon}$ (eV) | 2.9[e] | n.d. | 2.9[e] | [b](Su *et al.*, 2007) |
| $\beta$ | 0.41[e,c] | n.d. | 0.41[c] | [c](Robbins, 1980) |
| S | n.d. | n.d. | 0.61[c] | [d](Uehara *et al.*, 1960) |
| Q | 0.114[a] | 0.058[a] | 0.76-0.7[d] | [e](Lee *et al.*, 2016) |
| $\eta$ | n.d. | n.d. | 0.045[c] | |

Different efficiencies and properties of the $CaWO_4$ material at varying length scales. η is the overall efficiency for a process as defined in equation 10.

Given the fact that $CaWO_4$ NPs have UV emission characteristics motivated us to further explore the UV irradiance from $CaWO_4$ particles, and to assess what factors could possibly affect this emission. This would be useful in designing experiments to verify if UV emission is indeed the mechanism behind a $CaWO_4$ cancer killing effect. Two equations were derived to further our intuition into the problem, see SI material sections 3 and 4 for derivation details. One gives the effective flux through a nanoparticle, equation 8, and the other is a planar approximation to determine the flux on the surface of a detector measuring UV emission from the NPs, equation 9:

$$eqn\ (8) \quad f_{ph,i}^{NP} = \frac{N_{ph,i}}{4\pi R_p^2} = effective\ UV\ photon\ flux$$

$$\text{where } N_{ph,i} = \frac{(2\pi R_p^2)\hat{n}N_{ph}^o}{\alpha}e^{-\alpha R_p} + N_{ph}^o;$$

$$eqn\ (9) \quad f_{ph}^{detector} = \frac{\hat{n}N_{ph}^o}{2\alpha} = UV\ photon\ flux\ at\ a\ detector\ per\ history$$

where $N_{ph,i}$ is the total number of photons effectively passing through the surface of particle $i$, $\hat{n}$ is the NP number density in solution, $N_{ph}^o$ is the number of UV photons generated per NP per incident γ photon (eV/hist./NP), $\alpha$ is the absorbance at the UV photon wavelength, and $R_p$ is the radius of the NPs. The only unknown is the number of UV photons generated per NP per incident history, $N_{ph}^o$.

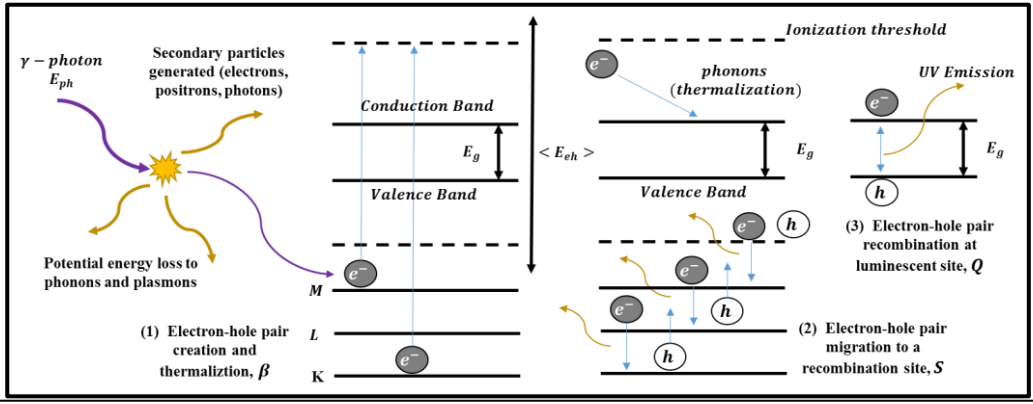

**Figure 7:** Simplified schematic of energy transfer from high energy radiation to luminescent centers – the scintillation process. The process is divided into roughly three stages, each represented by an efficiency parameter, equation 10: (1) Electron-hole pair creation and thermalization through plasmon interactions, $\beta$; (2) Electron-hole migration to a recombination center and thermalization through phonon interactions, $S$; (3) Electron-hole pair recombination at luminescent center, $Q$. Values for the varying efficiency are given in Table 1. Of the energy absorbed, the average energy to generate an electron-hole pair is $\langle E_{eh} \rangle$ in schematic or more precisely $2.35 \times E_g$ in equation 10. The ionization threshold is the energy value below which ionization cannot occur.

$N_{ph}^o$ can be calculated using theories first laid out by Roosbroeck and Robbins (van Roosbroeck, 1965; Robbins, 1980), which have been comprehensively discussed in the following (Mikhailik and Kraus, 2010; Nikl *et al.*, 2000; Rodnyi *et al.*, 1995; Lempicki *et al.*, 1993). The first equation, equation 10, is to determine the number of photons generated from the energy intrinsically absorbed by a NP, corresponds to the schematic in Figure 7:

$$eqn\ (10) \quad N_{ph}^o = \frac{E_\gamma^{abs}}{2.35 E_g} \beta S Q = N_{ph}^{max} \eta.$$

The energy absorbed, $E_\gamma^{abs}$, comes from simulation, and the 2.35 times the energy of the bandgap of the material ($E_g \approx 5.2$ eV for $CaWO_4$ (Lee *et al.*, 2016)) comes from the "crazy-carpentry" Monte Carlo simulations conducted by Roosbroeck (van Roosbroeck, 1965). The $\beta$, $S$ and $Q$ are all efficiencies for different processes in the energy conversion steps; $\beta$ can be calculated from material properties and represents energy loss to phonons and plasmons in the crystalline material, $S$ accounts for transmission losses as electrons and holes propagate through the material to a recombination site, and $Q$ is the quantum efficiency of an e-h pair at the recombination site. A summary of these values can be found in Table 1.

In both equations for either the effective photon flux through a nanoparticle surface or at the surface of a detector, equations 8 and 9, depend on the quantity $\hat{n} N_{ph}^o$. The number density of NPs, $\hat{n}$, and the number of photons generated per NP per incident γ photon (same as per history), $N_{ph}^o$. Through equation 10, it is apparent that $N_{ph}^o$ is proportional to $E_\gamma^{abs}$, the energy absorbed per NP. At a given mass density, the NP number density can be calculated as:

$$eqn\ (11) \quad \hat{n} = \frac{c_m}{\frac{4}{3}\pi R_p^3 \rho_p}$$

where $c_m$ is the total mass concentration of the NPs. And the specific energy absorbed has been shown to be independent of NP size, Figure 4(a), and therefore the $N_{ph}^o$ is proportional to:

$$eqn\ (12) \quad N_{ph}^o \propto E_\gamma^{abs} = \hat{E}_\gamma^{abs} * \frac{4}{3}\pi R_p^3 \rho_p$$

where, $\hat{E}_\gamma^{abs}$ is the specific energy absorbed per NP mass per history.

Therefore, it can be concluded that the overall intensity at a given mass concentration of CaWO$_4$ NPs is independent of NP size, because:

$$eqn\ (13) \quad \hat{n} N_{ph}^o \propto \frac{c_m}{\frac{4}{3}\pi R_p^3 \rho_p} * \hat{E}_\gamma^{abs} * \frac{4}{3}\pi R_p^3 \rho_p = c_m \hat{E}_\gamma^{abs}.$$

This result suggests that NP size does not affect the UV flux, indicating that size does not need to be optimized from the perspective of UV intensity. If taken from simulation, the results are on a per $\gamma$ photon incident (per history) basis. One could have conducted the derivation excluding the per $\gamma$ photon basis and reached the same conclusion. The number of impinging $\gamma$ photons (histories) per NP does change with NP size, and is determined by the cross-sectional area of the NP and the incident $\gamma$ photon flux, $f_{\gamma\ photon}$:

$$eqn\ (14) \quad histories\ per\ NP = \pi R_p^2 * f_{\gamma\ photon}.$$

The number of histories per NP varies proportionally to the $R_p^2$. Though this is not the end result, because a NP is comprised of absorbing centers that individually will receive the same number of incident $\gamma$ photons regardless of the size of the absorbing center agglomeration, i.e. the NP. To conclude, the intensity around a NP may vary as more absorbing centers are concentrated in a NP of increasing size, but macroscopically there is no difference in the number of UV photons generated. This analysis does not account for possible changes in the efficiencies of equation 10 with NP size, nor effects due to the increased proximity of an absorbing center to another absorbing center in a NP. The approximate order of magnitude of UV flux at 160 keV and 6 MeV $\gamma$ photon irradiation for a CaWO$_4$ NP in vacuum are approximately $1.0 \times 10^7$ and $1.0 \times 10^5$ photons per history per cm$^2$ of the NP surface; this is an effective quantity incorporating the incident UV photons from surrounding particles as well as photons generated from inside the particle. Future studies would need translate the number of UV photons emitted to a number of lethal lesions generated within a cell, in a form analogous to equation 6. It is the number of lethal lesions generated by a number of emitted electrons/photons that is logarithmically related to cell survival curves, not the energy of the electrons/photons, which is often represented through the linear-quadratic model as a function of dose either macroscopically or locally (McMahon *et al.*, 2011; Brenner, 2008); it should be noted that a particle emitted with a higher energy has the potential to generate more lethal lesions versus a particle of lesser energy.

## 5. Conclusion

The study has indicated that $HfO_2$ likely functions as Au NPs do for radio therapy enhancement. On the other hand, $CaWO_4$ has a DEF lower than that of Au and $HfO_2$, and as such is expected to function through a different mechanism or in combination with another mechanism. The conclusions were drawn from the local DEFs and relative enhancement ratios (RERs) between Au and $HfO_2$ or $CaWO_4$. To investigate what other mechanism by which $CaWO_4$ could enhance radiation therapy, its UV emission intensity was investigated. From this analysis it seems probable, or at this point unable to be ruled out, that $CaWO_4$ can impose damage through its UV emission. Theory and experiments indicate that $CaWO_4$ emits cell-toxic UV light. A significant amount of work utilizing carefully designed experiments will be necessary to fully elucidate $CaWO_4$'s functioning, but our findings indicate that $CaWO_4$ functions differently than Au and $HfO_2$. In closing, a word of caution is due to note the difficulty in translating simulations and theory to complex biological processes, but rather they help to develop our intuition into matters of great complexity, such as radio therapy enhancement.

**Acknowledgement**

Financial support for this work was derived from the Elevate Purdue Foundry Fund (EPFF) First-Tier Black Award funding awarded from the State of Indiana to LoDos Theranostics.

Supplemental Information (SI) for

# A Simulation Study on the Feasibility of Radio Enhancement Therapy with Calcium Tungstate and Hafnium Oxide Nanoparticles

Nicholas J. Sherck,[1,3] and You-Yeon Won[1,2,3]*

[1] *School of Chemical Engineering, Purdue University, West Lafayette, Indiana 47907*
[2] *Purdue University Center for Cancer Research, West Lafayette, Indiana 47907*
[3] *LoDos Theranostics, West Lafayette, Indiana 47906*

---

[*] To whom correspondence should be addressed. E-mail: yywon@ecn.purdue.edu

**Contents**

## Section 1: Count averages for electrons and photons

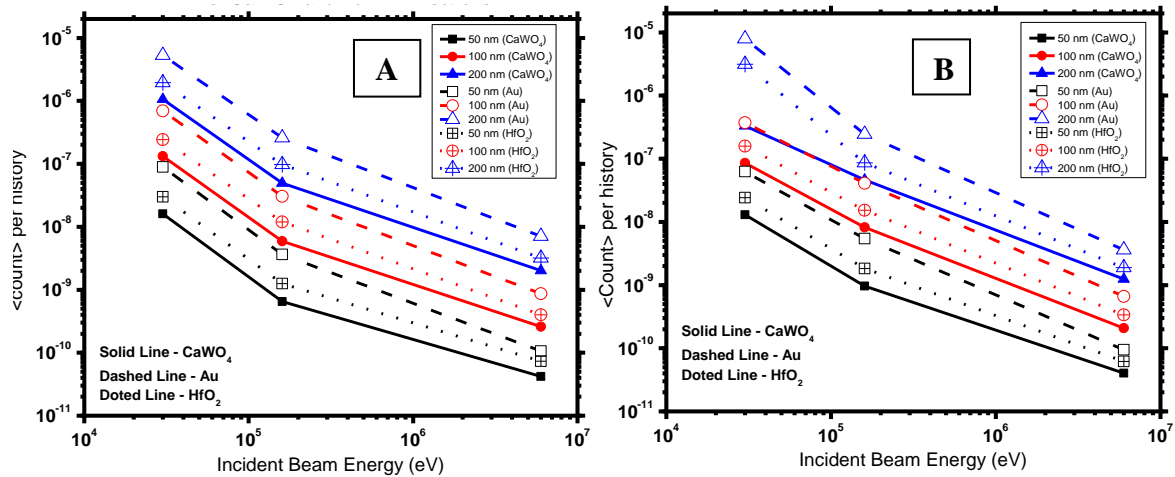

**SI Figure 1**: The count expectation value for Au, $CaWO_4$ and $HfO_2$ nanoparticles of varying sizes and incident monotonic, photon beam energies. (a) Electrons. (b) Photons. The count expectation value is calculated in accordance with equation 2. In general, from simulation Au exhibits a greater count per history and per nanoparticle size relative to both $CaWO_4$ and $HfO_2$. The count is expected to decrease by several orders of magnitude for each material with increasing incident beam energy due to the decrease in absorption by an isolated nanoparticle (increasing mean-free-paths for high energy particles). The counts are not normalized to particle mass, and when they are the particle count collapses for a given material and energy; this was not displayed here to maintain clarity in the image.

# Section 2: Local Dose Enhancement Ratio (DEF$_{loc}$) and Relative Enhancement Ratio (RER)

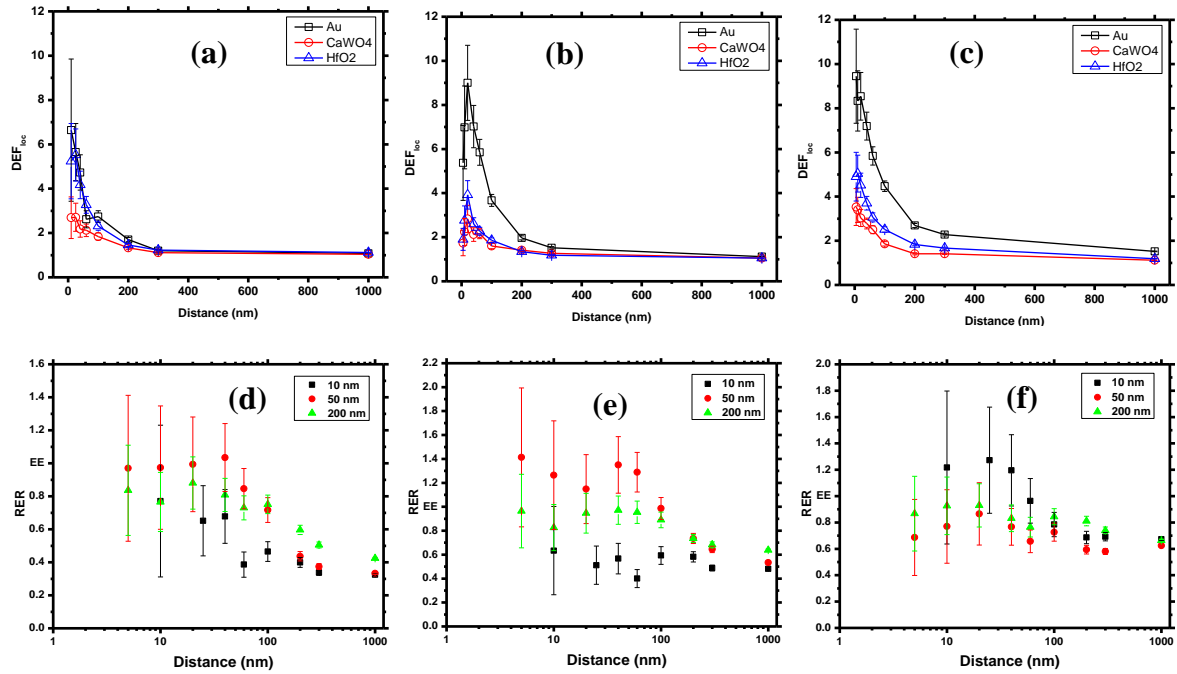

**SI Figure 2: (Top Row)** Plots of the local dose enhancement factor (DEF$_{loc}$) calculated at different distances from the surface of the NP (x-axis) for a monotonic, 6 MeV photon beam. (a) 10 nm NP. (b) 50 nm NP. (c) 200 nm NP. The error bars are ± 1 standard deviation. Simulations for each material against a baseline water medium under similar irradiation. **(Bottom Row)** Plots of the RER for (d) Au:CaWO$_4$, (e) Au:HfO$_2$, (f) HfO2:CaWO4 at varying distances from the surface.

# Section 3: Flux through a NP surface with multiple NPs

When given a number of NP emitting photons, even low energy, the MFP is quite high in an aqueous medium and continues for centimeters. It is known that in CaWO$_4$ themselves have a very low absorption coefficient of 0.231 cm$^{-1}$ and an absorption coefficient after annealing of 0.036 cm$^{-1}$ (Sivers *et al.*, 2012). Therefore, the absorption in nm sized particles is incredibly low and can be assumed negligible in this analysis.

The objective here is to determine the flux through a single nanoparticle given a number density of surrounding NPs emitting photons as well to get a cumulative flux. The idea is to integrate over the surface of a particle surrounded by NPs, radial number density *n(r)*, integrating spherically outward from the surface, $R_p$, to infinity:

$$N_{ph,i} = \int_{R_p}^{\infty} \int_0^{\pi} \int_0^{\pi} n(r) \frac{N_{ph}^o}{4\pi r^2} e^{-\alpha r} R_p^2 \sin(\theta) \, d\varphi d\theta dr$$

where

$N_{ph}^o = number\ of\ photons\ generated\ per\ particle,$

$\alpha = absorption\ coefficient\ of\ photons\ at\ a\ specific\ \lambda\ surrounding\ the\ particles,$

$$n(r) = \frac{dN}{dr} = 4\pi r^2 \hat{n}, \text{where } \hat{n} \text{ is the NP \# density.}$$

The bounds of integration only run over half the surface, given that a particle at a given distance from the NP of interest only sees half the NP of interest's surface. Upon integration one arrives at the equation:

$$N_{ph,i} = \frac{(2\pi R_p^2)\hat{n} N_{ph}^o}{\alpha} e^{-\alpha R_p} + N_{ph}^o.$$

And the flux through NP $i$ is given by:

$$f_{ph,i} = \frac{N_{ph,i}}{4\pi R_p^2}.$$

It is clear that the cumulative flux through the surface of any other nanoparticle depends on the number of photons generated by a NP, which depends on many factors, and the number density of the NPs.

**Section 4: Photon intensity at a detector – planar approximation**

Analogously to section 2, an experimentalist would also be interested in the flux of photons emitted by nanoparticles under irradiation, which is directly proportional to the intensity (flux per time). This equation makes use of a disk geometry a given distance from a point on the detector. By integrating outward from the detector, defined below as the z dimension, and while increasing the size of the disk radius, defined as y below, one can determine the flux at a given point on a planar surface:

$$f_{ph}^{detector} = \int_0^\infty \int_0^\pi \hat{n} f_{ph}^o(R) e^{-\alpha R} 2\pi y \, dy \, dz$$

where

$f_{ph}^o(R) = \#\ of\ ph.\ gen.\ per\ particle\ per\ surface\ area\ at\ the\ dist.\ of\ the\ detector,$

$\hat{n} = number\ density\ of\ NPs\ in\ the\ medium,$

$\alpha = absorption\ coefficient\ of\ photons\ at\ a\ specific\ \lambda\ surrounding\ the\ particles,$

$R^2 = y^2 + z^2.$

Given the exponential with the R term, which depends on both y and z, this equation is very difficult to handle. Thus, changing to spherical coordinates yields for a much more manageable integrand:

$$f_{ph}^{detector} = \int_0^\infty \hat{n} f_{ph}^o(R) e^{-\alpha R} 2\pi R^2 \, dR = \frac{1}{2} \int_0^\infty \hat{n} N_{ph}^o e^{-\alpha R} \, dR;$$

$$f_{ph}^{detector} = \frac{\hat{n} N_{ph}^o}{2\alpha}.$$

The equation for the flux of particles at a planar detector is very similar to that of the flux through any one NP distributed in the bulk of the medium.